\documentclass[12pt]{article}
\begin{document}

\title{\bf Linear Waves in Imperfect Relativistic MHD Fluid } 
\author{Sushil Kumar Singh and Subhash Kumar\thanks{E--mail: subhashkumar@rediffmail.com}
 \\ 
{\sl Department of Physics and Astrophysics,} \\
{\sl University of Delhi, Delhi-110007 (India)}} 

\date{}
\maketitle                   	
\baselineskip 22 pt

\begin{abstract}
\noindent Linear relativistic anisotropic magnetohydrodynamic waves are investigated for an Imperfect fluid. Consideration is carried out for the plasma with finite bulk viscosity. Conditions for growth and decay of these waves are obtained in special cases of isotropy and Chew-Goldberger-Low (CGL) state. 
\\
\\
{\small\bf{PACS number(s):51.20+d; 52.25.Xz;52.27.Ny;52.35.Bj}}
\end{abstract}

\clearpage
\vspace{0.7in}

\begin{section}*{\bf I. INTRODUCTION}
\noindent Relativistic plasma embedded in a strong magnetic field exhibits itself in a variety of interesting astrophysical events from pulsar winds to blackhole magnetospheres \cite{app, G&C, rees}. Waves and instabilities for a single fluid with isotropic pressure have been investigated in the relativistic magnetohydrodynamic framework \cite{hoff,lich}. The equation of state of plasma have also been subject of study when the strong magnetic field suppresses the equilibration of pressures parallel and perpendicular to itself \cite{gedeos}. Plane magnetosonic waves have been studied for a single relativistic fluid with anisotropic pressures \cite{ged, tshi}. However, systems in which radiation can provide the mechanism for viscosity and heat conduction, the bulk shear can be effective. Fluid (imperfect fluids) systems with isotropic pressure and dissipative effects have been put to investigations by Wienberg \cite{G&C, steve}.\\
\noindent The present paper intends to consider the effect of bulk viscosity on the plane wave propagation in relativistic magnetized plasma. The paper is organized as follows. In Sec. II we present the basic equations used in the present paper emphasizing the new energy-momentum tensor. In Sec. III the general dispersion relations are obtained and their derivation method is described briefly. In Sec. IV, to the first order of dissipation, the dispersion relation is discussed under various limits.             
\end{section}

\begin{section}*{\bf II. BASIC EQUATIONS}
\noindent We consider a plasma embedded in a magnetic field. Plasma fluid velocity and temperature both are allowed to be relativistic. Hereafter, we will use units where velocity of light $c=1$. Let $U^\alpha$ be the plasma 4-velocity and $F^{\alpha\beta}$ be the electromagnetic field tensor. The plasma rest-frame electric and magnetic fields are defined \cite{ged} as 
 
\begin{equation}
E^\alpha{\equiv}F^{\alpha\beta}U_\beta, 
\end{equation}

\begin{equation}
B^\alpha{\equiv}\frac{1}{2}\epsilon^{\alpha\beta\gamma\delta}U_{\beta}F_{\gamma\delta}, 
\end{equation}
respectively. The electromagnetic field tensor can then be written in the following form:
\begin{equation}
F^{\alpha\beta}=(E^{\alpha}U^\beta-E^{\beta}U^\alpha)+\frac{1}{2}\epsilon^{\alpha\beta\gamma\delta}(B_{\gamma}U_\delta-B_{\delta}U_\gamma),  
\end{equation} 
\noindent where $\epsilon^{\alpha\beta\gamma\delta}$ is a completely antisymmetric tensor and $\epsilon^{0123}=1$.

\noindent For a hydrodynamical system whose pressure, density and velocity vary appreciably over distances of the order of mean free path, thermal equibrium is not strictly maintained and the fluid kinetic energy is dissipated as heat. The presence of weak space-time gradients in an imperfect fluid has the effect of modifying the energy-momentum tensor \cite{steve}. For a fluid system interacting with radiation and wherein bulk viscosity is the dominant dissipative component, the stress energy-momentum tensor is  
\begin{equation}
T^{\alpha\beta}=W_{1}U^{\alpha}U^\beta-W_{2}\eta^{\alpha\beta}-W_{3}n^{\alpha}n^\beta-{\zeta}U^\gamma_{,\gamma}({U^\alpha}{U^\beta}-\eta^{\alpha\beta}),
\end{equation}

\noindent where
\begin{equation}
W_1=\varepsilon+p_\perp+\frac{B^2}{4\pi},
\end{equation}

\begin{equation}
W_2=p_\perp+\frac{B^2}{8\pi},
\end{equation}

\begin{equation}
W_3=p_\perp-p_\parallel+\frac{B^2}{4\pi}.
\end{equation}

\noindent Here $B^\alpha=Bn^\alpha$ and the coefficient of bulk viscosity $\zeta$ is positive definite. The plasma 4-velocity is normalized as $U^{\alpha}U_\alpha=1$. The magnetic field unit vector $n^\alpha$ obeys $n^{\alpha}n_\alpha=-1$. The energy density $\varepsilon$ is in general a function of mass density $\rho$ and the magnetic field strenght $B$, and can be put in the following functional form \cite{ged}:
\begin{equation}
\varepsilon={\rho}e(\rho,B)
\end{equation} 

\noindent The pressures are

\begin{equation}
p_\parallel=\rho^2\frac{\partial{e}}{\partial\rho}=\frac{\partial\varepsilon}{\partial{ln\rho}}-\varepsilon,
\end{equation}

\begin{equation}
p_\perp=p_\parallel+\rho{B}\frac{\partial{e}}{\partial{B}}=\frac{\partial\varepsilon}{\partial{ln\rho}}-\frac{\partial\varepsilon}{\partial{{ln}B}}-\varepsilon.
\end{equation}

\noindent These equations are in their most general form and are representative of relativistic anisotropic state.

\noindent Invoking the usual frozen-in condition $E^\alpha=0$ for the relativistic magnetohydrodynamic fluid, the electromagnetic dual tensor becomes
\begin{equation}
G^{\alpha\beta}=B(n^{\alpha}U^\beta-n^{\beta}U^\alpha), 
\end{equation}
and the Maxwell equations yields
\begin{equation}
G_{,\beta}^{\alpha\beta}=0.
\end{equation}

\noindent The equations of motion of the fluid are contained in the particle and energy-momentum conservation laws \cite{G&C}:

\begin{equation}
J_{,\alpha}^\alpha=0,
\end{equation}

\begin{equation}
T_{,\beta}^{\alpha\beta}=0,
\end{equation}

\noindent where $J^\alpha(={\rho}U^\alpha)$ is the 4-current density. 

\end{section}

\begin{section}*{\bf III. DISPERSION RELATION}

\noindent The perturbed state is characterized by $\rho \rightarrow \rho+\delta\rho, B \rightarrow B+\delta{B}, U^\alpha \rightarrow {U^\alpha}+\delta{U^\alpha}, n^\alpha \rightarrow n^\alpha+\delta{n^\alpha}$, and plane wave solutions for the perturbations require that $\delta{\rho}, \delta{B}, \delta{U^\alpha}, \delta{n^\alpha} \propto \exp{(ik_\alpha{x^\alpha})}$. It should be noted that the condition $U^{\alpha}U_\alpha=1$ and $n^{\alpha}n_\alpha=-1$ yields the following constraints, $U^{\alpha}\delta{U_\alpha}=n^{\alpha}\delta{n_\alpha}=0$.\\
\noindent It is assumed that the coefficient of bulk viscosity $\zeta$ does not vary appreciably with variations in above system parameters ($\rho, B, U^\alpha$). 

\noindent Substituting in Eqs.(12)-(14) $\partial/\partial{x^\alpha}\rightarrow ik_\alpha$, one obtains the following form:

\begin{equation}
k_\alpha\delta{J_\alpha}=0,
\end{equation}

\begin{equation}
k_\alpha\delta{G}^{\alpha\beta}=0,
\end{equation}

\begin{equation}
k_{\alpha}{\delta}T^{\alpha\beta}=0.
\end{equation}

\noindent The frequency and parallel component of the wave vector of the plane waves are defined in the rest frame of the fluid as:

\begin{equation}
\omega=k_\alpha{U^\alpha},~k_\parallel=-k_\alpha{n^\alpha}
\end{equation}
respectively, and
\begin{equation}
k_\alpha{k^\alpha}=s^2=\omega^2-k^2=\omega^2-k_\parallel^2-k_\perp^2.
\end{equation}

\noindent The coefficients $W_i$ in the expression for the energy-momentum tensor will be function of number density $\rho$ and magnetic field strength $B$. Hence,
$$
\delta{W_i}=\frac{\partial{W_i}}{\partial\rho}{\delta\rho}+\frac{\partial{W_i}}{\partial{B}}{\delta{B}}.
$$

\noindent The four-vector defined as  $l^\alpha\equiv\epsilon^{\alpha\beta\mu\nu}k_\beta{U_\mu}{n_\nu}$ \cite{ged} will be orthogonal to the independent vectors $k^\alpha$, $U^\alpha$ and $n^\alpha$ i.e., $l^\alpha{U_\alpha}=l^\alpha{n_\alpha}=l^\alpha{k_\alpha}=0$. Multiplication of Eqs.(15)-(17) by $l_\alpha$ , one obtains

\begin{equation}
\left(
\begin{array}{c c}
k_\parallel & \omega \\
\omega{W_1} & k_\parallel{W_3}
\end{array}
\right)
\left(\begin{array}{c}
l_\alpha\delta{U^\alpha}\\
l_\alpha\delta{n^\alpha}
\end{array}
\right)=0.
\end{equation}

\noindent This yields the corresponding dispersion relation for the Alfven wave:
\begin{eqnarray}
\omega^2 &=& \frac{W_3}{W_1}k_\parallel^2, \nonumber \\
&=& \bigg(\frac{p_\perp-p_\parallel+\frac{B^2}{4\pi}}{\varepsilon+p_\perp+\frac{B^2}{4\pi}}\bigg)k_\parallel^2. 
\end{eqnarray}

\noindent Multiplying Eqs.(15)-(17) by the other three independent vectors  $k^\alpha$, $U^\alpha$ and $n^\alpha$ gives the following matrix

\begin{equation}
\left(
\begin{array}{c c c c c c}
\omega & 0 & 0 & 0 & 0 & 1 \\
0 & -k_\parallel & 1 & -\omega & 0 & 0 \\
0 & \omega & 0 & 0 & -k_\parallel & 1 \\
A_1 & B_1 & 0 & k_\parallel{W_3} & 0 & W_1 \\
A_2 & B_2 & 2k_\parallel{W_3} & 0 & 0 & -2i\zeta\omega{W_1}k^2 \\ 
A_3 & B_3 & W_3 & 0 & \omega{W_1} & -i\zeta{k_\parallel} 
\end{array}
\right)
\left(\begin{array}{c}
\delta{ln\rho}\\
\delta{ln{B}}\\
k_\alpha\delta{n^\alpha}\\
U_\alpha\delta{n^\alpha}\\
n_\alpha\delta{U^\alpha}\\
k_\alpha\delta{U^\alpha}\\
\end{array}
\right)=0,
\end{equation}

\noindent where 
$$
A_1=\omega\frac{\partial}{\partial{ln\rho}}(W_1-W_2),
$$

$$
A_3=k_\parallel\frac{\partial}{\partial{ln\rho}}(W_2-W_3),
$$

$$
A_2=\omega{A_1}+k_\parallel{A_3}+k_\perp^2\frac{\partial}{\partial{ln\rho}}W_2,
$$

$$
B_1=\omega\frac{\partial}{\partial{lnB}}(W_1-W_2),
$$

$$
B_3=k_\parallel\frac{\partial}{\partial{lnB}}(W_2-W_3),
$$

$$
B_2=\omega{B_1}+k_\parallel{B_3}+k_\perp^2\frac{\partial}{\partial{lnB}}W_2.
$$

\noindent The corresponding dispersion relation for the slow and fast magnetosonic modes are given by

\begin{eqnarray}
\bigg[\omega^2\frac{\partial(W_1-W_2)}{\partial{ln\rho}} - k_\parallel^2\frac{\partial(W_2-W_3)}{\partial{ln\rho}}\bigg]\bigg[k_\parallel^2{W_3}-\omega^2{W_1}+k_\perp^2\frac{\partial{W_2}}{\partial{lnB}}\bigg] \nonumber\\
+ k_\perp^2\frac{\partial{W_2}}{\partial{ln\rho}}\bigg[k_\parallel^2\bigg(W_3+\frac{\partial(W_2-W_3)}{\partial{lnB}}\bigg)+\omega^2\bigg(W_1-\frac{\partial(W_1-W_2)}{\partial{lnB}}\bigg)\bigg] \nonumber\\
-i\zeta\omega\bigg[k_\parallel^2\bigg(k_\parallel^2{W_3}-\omega^2{W_1}\bigg)+k_\perp^2\bigg(\omega^2\frac{\partial(W_1-W_2)}{\partial{lnB}}+k_\parallel^2\frac{\partial(W_3)}{\partial{lnB}}\bigg)\bigg]=0.
\end{eqnarray}

\noindent In the limit of bulk viscosity $\zeta\rightarrow0$ (perfect RAM fluid), this equation reduces to the dispersion relation obtained by Gedalin \cite{ged}.
\end{section}

\begin{section}*{\bf IV. DISCUSSION}

{\bf{Unmagnetized Plasma :}} \\ 
\noindent In the absence of magnetic field, the energy density $\varepsilon=\varepsilon{(\rho)}$ and isotropy prevails i.e., $W_3=0$; and to the first order of dissipative term \cite{steve}, we obtain 

\begin{equation}
\omega=\omega_0+iLk^2,
\end{equation}
where the decay characteristic length 
\begin{equation}
L=\frac{\zeta}{2(\varepsilon+p)},
\end{equation}
and the frequency of the sound waves with the bulk shear being neglected 
\begin{eqnarray}
\omega_0 &=& k \bigg\{\frac{\frac{\partial{W_2}}{\partial{ln\rho}}}{\frac{\partial}{\partial{ln\rho}}(W_1-W_2)} \bigg\}^{\frac{1}{2}},\nonumber\\
&=& k \bigg(\frac{\partial{p}}{\partial\varepsilon}\bigg)^{\frac{1}{2}}.
\end{eqnarray}
\noindent The amplitude of the sonic waves decays at the rate 
$\Gamma=Lk^2$.

\noindent {\bf{Magnetized Plasma :}}\\
\noindent (i) In the special case of propagation of magnetosonic modes in the direction of the magnetic field, the frequency is found to be
\begin{equation}
\omega=\omega_0+iL_\parallel{k^2},
\end{equation}
with the decay characteristic length 
\begin{equation}
L_\parallel=\frac{\zeta}{2(\varepsilon+p_\parallel)},
\end{equation}
and the viscosity neglected frequency 
\begin{eqnarray}
\omega_0 &=& k \bigg\{\frac{\frac{\partial{(W_2-W_3)}}{\partial{ln\rho}}}{\frac{\partial}{\partial{ln\rho}}(W_1-W_2)} \bigg\}^{\frac{1}{2}},\nonumber\\
&=& k \bigg\{\frac{\frac{\partial{p_\parallel}}{\partial{ln\rho}}}{(\varepsilon+p_\parallel)}\bigg\}^{\frac{1}{2}}.
\end{eqnarray}

\noindent It is to be noted that $L_\parallel$ is always positive. This implies that the wave decays for all possible values of the system parameters. 

\noindent (ii) For the magnetosonic waves propagating perpendicular to the ambient magnetic field, the frequency is
\begin{equation}
\omega=\omega_0-iL_\perp{k^2},
\end{equation}
where the spatial scale of decay 
\begin{equation}
L_\perp=\frac{\zeta(\frac{\partial\varepsilon}{\partial{lnB}}+\frac{B^2}{4\pi})}{2(\varepsilon+p_\perp+\frac{B^2}{4\pi})\frac{\partial\varepsilon}{\partial{ln\rho}}},
\end{equation}
and the frequency in the absence of dissipation 
\begin{eqnarray}
\omega_0 &=& k \bigg\{\frac{\frac{\partial{(W_1-W_2)}}{\partial{ln\rho}}\frac{\partial{W_2}}{\partial{lnB}}+\frac{\partial{W_2}}{\partial{ln\rho}}[W_1-\frac{\partial{(W_1-W_2)}}{\partial{lnB}}]}{W_1\frac{\partial(W_1-W_2)}{\partial{ln\rho}}} \bigg\}^{\frac{1}{2}},\nonumber\\
&=& k \bigg\{ \frac{\frac{\partial{p_\perp}}{\partial{lnB}}+\frac{B^2}{4\pi}+\frac{\frac{\partial{p_\perp}}{\partial{ln\rho}}}{\frac{\partial{\varepsilon}}{\partial{ln\rho}}}(\varepsilon+p_\perp-\frac{\partial{\varepsilon}}{\partial{lnB}})}{\varepsilon+p_\perp+\frac{B^2}{4\pi}} \bigg\}^{\frac{1}{2}}.
\end{eqnarray}

\noindent In this case we observe that the magnetosonic modes can grow or decay accordingly as $L_\perp>0$ or $L_\perp<0$. In the latter case the system tends to return to its equilibrium state while in the former instability grows. A state characterized by $L_\perp=0$ represents propagation of harmonics in the direction transverse to the magnetic field unlike the unmagnetized-isotropic state wherein the harmonics can propagate only upto its few characteristic lengths. \\   
\noindent Shrauner \cite{shr} proposed a generalized polytrope model for the MHD set of equations for considering the collisional, collisionless and the transitional regimes of plasma state. According to Shrauner, the anisotropic pressures can be written as

\begin{equation}
p_\parallel=C_\parallel\rho^\nu{B^{-\alpha}},~~~p_\perp=C_\perp\rho^\delta{B^{\beta}}, 
\end{equation}

\noindent where $C_\parallel$ and $C_\perp$ are positive constants. The polytropic indices $\nu, \alpha, \delta$ and $\beta$ are positive constants and these polytropic relations can be reduced in certain special cases to well known equations of state for the pressure.  
From equations (8), (9), (10) and (33) we get the generalized energy density 
\begin{equation}
\varepsilon=a_1\rho+a_2{p_\parallel}+a_3{p_\perp}, 
\end{equation}

\noindent where $a_1, a_2$ and $a_3$ are positive constants. The condition for decay of the transverse modes i.e., $L_\perp<0$ translate into
\begin{equation}
a_2\alpha{p_\parallel}>a_3\beta{p_\perp}+\frac{B^2}{4\pi}.
\end{equation}
\\
\noindent For $a_1=1, a_2=1/2$ and $a_3=1$ we recover the energy density for a plasma state characterized by nonrelativistic temperatures; $ \varepsilon=\rho+\frac{1}{2}{p_\parallel}+{p_\perp} $. Further, if $\nu=3, \alpha=2, \delta=1$ and $\beta=1$, then the system represents the well known CGL \cite{cgl} state with $p_\parallel=C_\parallel\rho^3/{B^2}$ and $p_\perp=C_\perp\rho{B}$. It is evident that the condition for decay transverse modes in the CGL plasma reads as

\begin{equation}
p_\parallel>p_\perp+\frac{B^2}{4\pi}.
\end{equation}

\noindent This is the usual criterion for firehose instability in MHD.

\noindent We would like to point out that a simple gas of structureless point particles will have negligible bulk viscosity in the extreme-relativistic or nonrelativistic limits. However, this need not be the case for a fluid composed of a mixture of highly relativistic and nonrelativistic particles. It is well known that the exchange of energy between translational and rotational degrees of freedom gives ordinary diatomic gases an appreciable bulk viscosity. Although the present analysis assumes that coefficient of bulk viscosity $\zeta$ does not vary appreciably, it does depend on the dynamical variables of the system in the general form $\zeta=4bT^4\tau[\frac{1}{3}-(\frac{\partial{p}}{\partial\varepsilon})_{\rho}]^2$. Here T is plasma temperature, b is of the order of Stefan-Boltzmann constant and $\tau$ is the free mean time of radiation quanta \cite{steve}. In view of this, the present work is an indicative analysis of plasma as an imperfect fluid and can be explored further.

\end{section}

\end{document}